\documentclass[12pt]{iopart}
\usepackage{graphicx}
\usepackage{epsfig}
\usepackage{citesort}
\usepackage{xspace}

\usepackage{iopams}
\usepackage{color}

\newcommand{\text}{\rm }
\newcommand{\hefour}{\ensuremath{{}^{4}{\rm He}}\xspace}
\newcommand{\hethree}{\ensuremath{{}^{3}{\rm He}}\xspace}

\newcommand{\gag}{\ensuremath{g_{{\rm a}\gamma}}}
\newcommand{\ma}{\ensuremath{{m_{\rm a}}}}

\begin{document}
\title[]{Probing eV-scale axions with CAST}
\author{E.~Arik$^{1,\dagger}$,
S.~Aune$^{2}$,
D.~Autiero$^{3,18}$, K.~Barth$^{3}$, A.~Belov$^{4}$,
B.~Beltr\'an$^{5,19}$, S.~Borghi$^{3,20}$, G. Bourlis$^{21}$,
F.~S.~Boydag$^{1,\dagger}$,
H.~Br\"auninger$^{6}$, J.~M.~Carmona$^{5}$, S.~Cebri\'an$^{5}$,
S.~A.~Cetin$^{1}$, J.~I.~Collar$^{7}$, T.~Dafni$^{5}$,
M.~Davenport$^{3}$, L.~Di~Lella$^{3,22}$,
O.~B.~Dogan$^{1,\dagger}$,
C.~Eleftheriadis$^{8}$, N.~Elias$^{3}$,
G.~Fanourakis$^{9}$, E.~Ferrer-Ribas$^{2}$, H.~Fischer$^{10}$,
P.~Friedrich$^{6}$,
J.~Franz$^{10}$, J.~Gal\'an$^{5}$,
T.~Geralis$^{9}$, I.~Giomataris$^{2}$, S.~Gninenko$^{4}$,
H.~G\'omez$^{5}$, 
R.~Hartmann$^{6,23}$,
M.~Hasinoff$^{11}$, F.~H.~Heinsius$^{10,24}$,
I.~Hikmet$^{1,\dagger}$,
D.~H.~H.~Hoffmann$^{12}$,
I.~G.~Irastorza$^{5}$, J.~Jacoby$^{13}$,
K.~Jakov\v{c}i\'{c}$^{14}$, D.~Kang$^{10,25}$, K.~K\"onigsmann$^{10}$,
R.~Kotthaus$^{15}$, M.~Kr\v{c}mar$^{14}$, K.~Kousouris$^{9,26}$,
M.~Kuster$^{6,12}$, B.~Laki\'{c}$^{14}$, C.~Lasseur$^{3}$,
A.~Liolios$^{8}$, A.~Ljubi\v{c}i\'{c}$^{14}$,
G.~Lutz$^{14}$, G.~Luz\'on$^{5}$, D.~Miller$^{7}$,
J.~Morales$^{5}$,
T.~Niinikoski$^{3}$, A.~Nordt$^{6,12}$, A.~Ortiz$^{5}$,
T.~Papaevangelou$^{2}$, M.~J.~Pivovaroff$^{16}$, A.~Placci$^{3}$,
 G.~Raffelt$^{15}$, H.~Riege$^{3,12}$,
A.~Rodr\'iguez$^{5}$, J.~Ruz$^{5}$, I.~Savvidis$^{8}$,
Y.~Semertzidis$^{17,27}$,
P.~Serpico$^{15,28}$,
R.~Soufli$^{16}$, L.~Stewart$^{3}$,
K.~van~Bibber$^{16}$, J.~Villar$^{5}$, J.~Vogel$^{10}$,
L.~Walckiers$^{3}$, K.~Zioutas$^{17,3}$(CAST Collaboration)}

\address{$^1$ Dogus University, Istanbul, Turkey}
\address{$^2$ CEA, IRFU, Centre de Saclay, Gif-sur-Yvette, France}
\address{$^3$ European Organization for Nuclear Research (CERN), Gen\`eve, Switzerland}
\address{$^4$ Institute for Nuclear Research (INR), Russian Academy of Sciences, Moscow, Russia}
\address{$^5$ Instituto de F\'{\i}sica Nuclear y Altas Energ\'{\i}as, Universidad de Zaragoza, Zaragoza, Spain}
\address{$^6$ Max-Planck-Institut f\"{u}r extraterrestrische Physik, Garching, Germany}
\address{$^7$ Enrico Fermi Institute and KICP, University of Chicago, Chicago, IL, USA}
\address{$^8$ Aristotle University of Thessaloniki, Thessaloniki, Greece}
\address{$^9$ National Center for Scientific Research ``Demokritos'', Athens, Greece}
\address{$^{10}$ Albert-Ludwigs-Universit\"{a}t Freiburg, Freiburg, Germany}
\address{$^{11}$ Department of Physics and Astronomy, University of British Columbia, Vancouver, Canada}
\address{$^{12}$ Technische Universit\"{a}t Darmstadt, IKP, Darmstadt, Germany}
\address{$^{13}$ Johann Wolfgang Goethe-Universit\"at, Institut f\"ur Angewandte Physik, Frankfurt am Main, Germany}
\address{$^{14}$ Rudjer Bo\v{s}kovi\'{c} Institute, Zagreb, Croatia}
\address{$^{15}$ Max-Planck-Institut f\"{u}r Physik, M\"unchen, Germany}
\address{$^{16}$ Lawrence Livermore National Laboratory, Livermore, CA, USA}
\address{$^{17}$ Physics Department, University of Patras, Patras, Greece}

\vspace{1cm}

\address{$^{18}$ Present address: Institut de Physique Nucl\'eaire, Lyon, France}
\address{$^{19}$ Present address: Department of Physics, University of Alberta, Edmonton, Alberta, Canada}
\address{$^{20}$ Present address: Department of Physics and Astronomy, University of Glasgow, Glasgow, UK}
\address{$^{21}$ Present address: Hellenic Open University, Patras, Greece}
\address{$^{22}$ Present address: Scuola Normale Superiore, Pisa, Italy}
\address{$^{23}$ Present address: PNSensor GmbH, Römerstrasse 28, M\"unchen, Germany}
\address{$^{24}$ Present address: Ruhr-Universit\"{a}t Bochum, Bochum, Germany}
\address{$^{25}$ Present address: Institut f\"ur Experimentelle Kernphysik, Universit\"at Karlsruhe, Karlsruhe, Germany}
\address{$^{26}$ Present address: Fermi National Accelerator Laboratory, Batavia, Illinois, USA}
\address{$^{27}$ Present address: Brookhaven National Laboratory, Upton, New York, USA}
\address{$^{28}$ Present address: European Organization for Nuclear Research (CERN), Gen\`eve, Switzerland }
\address{$^{\dagger}$ Deceased}
\ead{eferrer@cea.fr}
\begin{abstract}

We have searched for solar axions or other pseudoscalar particles that couple
to two photons by using the CERN Axion Solar Telescope (CAST) setup.
Whereas we previously have reported results from CAST with evacuated magnet bores
(Phase I), setting limits on lower mass axions, here we report results
from CAST where the magnet bores were filled with \hefour gas (Phase II) of variable pressure.
The introduction of gas generates a refractive photon mass $m_\gamma$, thereby achieving the maximum possible
conversion rate for those axion masses \ma~that match $m_\gamma$. 
With 160 different pressure settings we have scanned
\ma~up to about 0.4\;eV, taking approximately 2\;h of data for
each setting. From the absence of excess X-rays when the magnet was
pointing to the Sun, we set a typical upper limit on the
axion-photon coupling of $\gag\lesssim 2.2\times
10^{-10}\;{\rm GeV}^{-1}$ at 95\% CL for $\ma \lesssim
0.4$\;eV, the exact result depending on the pressure setting. The
excluded parameter range covers realistic axion models with a
Peccei-Quinn scale in the neighborhood of $f_{\rm a}\sim10^{7}$\;GeV.
Currently in the second part of CAST Phase II, we are searching for axions with 
masses up to about 1.2\;eV using \hethree as a buffer gas.

\end{abstract}
\pacs{95.35.+d; 14.80.Mz; 07.85.Nc; 84.71.Ba}
\maketitle

\newpage

\section{Introduction}                        \label{sec:introduction}

The most promising experimental approach to search for axions or other
pseudoscalar particles is to use their two-photon interaction that is
traditionally written in the form
\begin{equation}\label{eq:axionphotoncoupling}
     {\cal L}_{{\rm a}\gamma}=
     -\frac{1}{4}\,g_{{\rm a}\gamma} F_{\mu\nu}\tilde F^{\mu\nu}a
     =g_{{\rm a}\gamma}\,{\bf E}\cdot{\bf B}\,a\,,
\end{equation}
where $a$ is the axion field, $F$ the electromagnetic field-strength
tensor, $\tilde F$ its dual, ${\bf E}$ the electric, ${\bf B}$ the magnetic
field, and \gag~the axion-photon coupling constant. This
interaction implies the conversion $a\leftrightarrow\gamma$ in the
presence of external electric and magnetic fields. One strategy 
(see \cite{Battesti:2007um} for a review) to search
for this effect is Sikivie's helioscope technique where a long dipole
magnet is oriented towards the Sun~\cite{Sikivie:1983ip}. Axions with
energies of a few~keV would be produced in the hot solar interior by the
transformation of thermal photons into axions in the electric
fields of the charged particles of the solar plasma, the Primakoff
effect~\cite{Dicus:1978fp,Raffelt:1985nk}, and re-converted into x-rays
within the dipole magnet. This conversion in a macroscopic $B$ field is
best viewed as a particle oscillation phenomenon in analogy to neutrino
flavor oscillations~\cite{Raffelt:1987im}.  Accordingly, the conversion
probability in a $B$-field region of length $L$ is~\cite{vanBibber:1988ge}
\begin{equation}\label{eq:conversionprobability}
      P_{{\rm a}\to\gamma}=\left( \frac{g_{{\rm a}\gamma}B}
          {q} \right)^2 \sin^2\left( \frac{qL}{2} \right)\,,
\end{equation}
where in vacuum $q=m_{\rm a}^2/2E$ is the photon-axion momentum
difference (we use natural units with $\hbar=c=1$).

An early helioscope experiment was performed in Brookhaven in the early
1990s~\cite{Lazarus:1992ry} and later a much more sensitive search in
Tokyo~\cite{Moriyama:1998kd,Inoue:2002qy,Inoue:2008}. The largest and most sensitive
helioscope, the CERN Axion Solar Telescope (CAST), has taken data since
2003 and has provided the most restrictive constraints on the axion-photon
coupling~\cite{Zio05,And07}, superseding well-known
astrophysical limits~\cite{Raffelt:2006cw}. 
The limit, $\gag\lesssim 8.8\times
10^{-11}\;{\rm GeV}^{-1}$, obtained in the first phase of CAST (CAST-I)\cite{And07} applies only for masses $\ma \lesssim 0.02$\;eV.
This is seen most easily by rewriting the conversion probability in the form

\begin{equation}\label{eq:conversionprobability2}
      P_{{\rm a}\to\gamma}=\left( \frac{g_{{\rm a}\gamma}B\,L}{2} \right)^2 \frac{\sin^2(x)}{x^2}\,,
\end{equation}
where $x\equiv qL/2$. The degradation of sensitivity to $g_{{\rm
    a}\gamma}$ when going to high masses follows then from the
$x^{-2}$ suppression of the conversion probability when $x\gg 1$.  For
the magnet length $L\simeq 9.26$~m and the typical 4~keV energy
of solar axions, the sensitivity declines when $x\gtrsim 1$, or
$m_{\rm a}\gtrsim \sqrt{4E/L}\simeq 0.02$~eV.


It has long been recognized that the sensitivity of axion helioscope
experiments can be extended to larger masses if one fills the
conversion region with a suitable buffer gas, providing the photons
with an effective mass $m_\gamma$~\cite{vanBibber:1988ge}. The
axion-photon momentum difference becomes $q=(m_{\rm
  a}^2-m_\gamma^2)/2E$ so that for \ma~values in the
neighborhood of the chosen $m_\gamma$ the maximum sensitivity is
restored. Varying the gas density allows one to scan an entire
range of $m_{\rm a}$ values, of course at the price of having to
take data separately at each density setting. Such a programme was
first carried out at the Tokyo axion helioscope~\cite{Inoue:2002qy} with recent results~\cite{Inoue:2008}. 
In 2005 CAST transitioned to its second phase (CAST-II), operating with a buffer gas
to increase sensitivity to higher axion masses.
In the first part of CAST-II from late 2005 to early 2007,
data were taken with $^4$He as a buffer gas, extending the
sensitivity to $m_{\rm a}\lesssim 0.4$\;eV. In this manuscript, we report the results of these measurements
that supersede all previous laboratory limits on the axion-photon
coupling strength in this mass range.

In general the dispersion relation of photons in matter is a complicated
function of energy. The refractive index can be either larger or smaller
than unity so that in the medium, the dispersion relation can be either space-like 
($E^2-p^2<0$) or time-like ($E^2-p^2>0$). In the former case, e.g.\
for visible light in air, the mismatch between the axion and photon
dispersion relations would be exacerbated. We stress this point because it
is sometimes overlooked in the literature. If the photon energy is far
above all resonances of the medium it is guaranteed, however, that the
dispersion relation is not only time-like, but also that its energy
dependence is such as if the photons had an effective mass:
$E^2-p^2=m_\gamma^2$ where $m_\gamma^2=\omega_{\rm plas}^2=4\pi\alpha
n_e/m_e$ with $n_e$ the electron density, $m_e$ the electron mass, and
$\alpha=1/137$ the fine-structure constant.  In other words, for
high-energy photons any medium has the same effect as a nonrelativistic
plasma with the same electron density.  For x-ray energies of a few keV
relevant for CAST, one needs to use a low-$Z$ gas such as hydrogen or
helium~\cite{vanBibber:1988ge}. In the first part of CAST-II we use $^4$He, allowing us to go
up to a pressure of 16.4\;mbar, the $^4$He vapor pressure at the operating
temperature of 1.8\;K of the superconducting magnet, corresponding to
$m_\gamma \approx 0.4$\;eV. To reach larger masses up to about 1.2\;eV we use
$^3$He in the second part of CAST-II. 

CAST and previous helioscope experiments are sensitive to axion-like
particles within a certain region in the two-parameter plane spanned
by $g_{{\rm a}\gamma}$ and $m_{\rm a}$. The main motivation,
however, is to search for QCD axions that appear as a consequence of
the Peccei-Quinn mechanism to solve the CP problem of strong
interactions~\cite{Peccei:2006as}. For these particles we have
\begin{eqnarray}
 g_{{\rm a}\gamma}&=&
 \frac{\alpha}{2 \pi}\, \frac{1}{f_{\rm a}}\,
 \left( \frac{E}{N} - \frac{2}{3}\,\frac{4+z}{1+z}\right)\,,
 \nonumber\\
 m_{{\rm a}}&=&
 \frac{z^{1/2}}{1+z}\, \frac{f_{\pi} m_{\pi}}{f_{\rm a}}\,,
\end{eqnarray}
where $f_{\rm a}$ is the Peccei-Quinn scale or axion decay constant,
a free parameter of the theory, $m_\pi$ is the pion mass,
$f_\pi=92$\;MeV the pion decay constant, $E/N$ the ratio of the
electromagnetic and colour anomalies of the axial current associated
with the Peccei-Quinn $U(1)$ symmetry, and $z=m_{\rm u}/m_{\rm d}$
the up/down quark mass ratio. Therefore, in the $g_{{\rm
a}\gamma}$--$m_{\rm a}$ plane, the locus for QCD axions is given by
\begin{equation}
 g_{{\rm a}\gamma}=
 \frac{\alpha}{2 \pi}\,
 \left( \frac{E}{N} - \frac{2}{3}\,\frac{4+z}{1+z}\right)
 \frac{1+z}{z^{1/2}}\,
 \frac{m_{\rm a}}{m_\pi f_\pi}\,,
\end{equation}
sometimes called ``the axion line.'' Since $E/N$ is a
model-dependent parameter of order unity and since $z$ is somewhat
uncertain~\cite{Yao:2006px}, QCD axions, if they exist, presumably
live somewhere within a relatively narrow band in the $g_{{\rm
a}\gamma}$--$m_{\rm a}$ plane shown as a yellow band in figure~\ref{fig:exclusion}.
While the CAST-I limits did not yet intersect the axion band, the
CAST-II results reported below begin to encroach into it.

The CAST magnet, the detectors, and the Sun as an axion source were
described in detail in the final CAST-I paper~\cite{And07} and~\cite{Kus07,Aut07,Abb07}. In
section~\ref{sec:experimental} we describe the modifications of the CAST experimental set-up necessary
to inject a buffer gas in the magnet pipes with the required precision, followed by a description of the strategy that we have followed for the data taking. We describe our data analysis and provide new limits in section~\ref{sec:analysis} before concluding in section~\ref{sec:conclusion}.

\section{Upgrade and strategy of the CAST experiment for Phase II}
\label{sec:experimental}
The setup of the CAST experiment has been described
elsewhere~\cite{Zio05,Zio99}. Here we focus on the upgrade that was done to allow operation with \hefour buffer gas in the cold magnet bore. This upgrade was the first part of a transition to a more complex system designed for eventual operation at higher buffer gas densities using \hethree. The system was designed after a series of measurements and tests during resistive transitions (quenches) of the superconducting magnet in June 2005. It was found that that the helium pressure in the cold bore rose by a factor 13 in the first 3 seconds after the quench trigger, and by a factor 20 in 120 seconds, when the gas in the cold bore tubes was not extracted.

During the \hefour run the density of the helium in the cold bores was increased in daily steps equivalent to a pressure step of 0.08\;mbar at 1.8\;K in the cold bores and in the pipework linking it to the gas system outside. In order to be able to reproduce the density settings precisely, the steps were controlled by injecting a precisely metered quantity of \hefour gas into the cold bores. The metering was based on a precise volume with temperature control within 0.1$^o$C, and on a metrology grade pressure gauge. The density step was determined so as to shift the peak in the axion mass acceptance by one FWHM that provides a sizeable overlap with the previous setting. Thus by making a series of steps, resulting in a rather smooth scan of the axion mass range, the discovery potential is maximized and has no substantial mass gaps. The system was designed to operate up to 16.4\;mbar pressure at 1.8\;K, with a stable and homogeneous density along the cold bores, and with an accuracy and reproducibility of equivalent pressure settings better than 0.01\;mbar and 0.1\;mbar, respectively~\cite{SPSCHe4,Tapio08,He3TDR}. The homogeneity of the density along the cold bores of the magnet is ensured by the efficient thermal coupling with the subcooled superfluid liquid helium bath filling the magnet vessel. Moreover, as the buffer gas is sealed within the volume including the cold bores and the dead volumes of the linking pipework, the density remains sufficiently constant despite of the small, uncorrelated fluctuations of the temperatures of the magnet and of the dead volumes, because care was taken to minimise the volume of the external pipework connected to the cold bore. 
During the initial tests spontaneous thermo-acoustic oscillations (TAOs) were observed in the pipework formed by the cold bores and their connections up to the room-temperature shut-off valves. The standing wave has 3.7 Hz fundamental frequency in our system and is driven by non-linear forces in the pipes with large temperature gradient. As a consequence the density varies in time and space in the cold bores, which results in a loss of sensitivity and control of the axion peak width. In order to avoid this effect, damping elements were installed inside the interconnecting pipework to entirely eliminate the conditions that generate the TAO. The tests during magnet quenches were required also for the design of the \hethree gas system components and of the passive and active safety measures to protect the cold X-ray windows against the pressure surges due to the quenches. This was not used during the operation with \hefour as buffer gas reported here, and will be the subject of a forthcoming paper.

The effect of the gravity in the gas density in the cold bores was also studied and was found to be negligible even if the cold bore tubes were operated vertically. The smallness of the maximum tilt angle of $\pm 8^o$ reduces the effect further.

A key element in the system are the x-ray windows that confine the buffer gas axially. Four windows were installed at the entrance and exit of the two cold bores of the magnet.
These windows must have a high x-ray transmission in the range of 2\;keV to 8\;keV. This requires the use of very thin low Z-material, such as beryllium or plastic. Additional requirements for these windows are:   i)resistance to static and dynamic pressures during normal operation and when the magnet quenches, ii)low permeability to helium and iii)transparency at visible wavelengths for both visual inspection and for laser alignment of the x-ray telescope and CCD relative to the axis of the cold bore.

In order to fulfill these requirements, we have made the windows of 15\;$\mu$m thick polypropylene film, supported by an electro-eroded grid (strongback) structure to withstand the 150\;mbar pressure difference of future normal operation with \hethree, as well as the rapid pressure rise that may happen during a quench of the magnet. The strongback obscures 12.6\% of the geometric area. Although polypropylene has a large helium permeation rate at room temperature, this rate is reduced by at least 4 orders of magnitude at cryogenic temperatures.  All windows were pressure tested at 1.5\;bar at cryogenic temperatures, and the results of leak tests showed values better than  $1\times 10^{-7}\mbox{\rm{mbar}}\mbox{\rm{L}}\mbox{\rm{s}}^{-1}$ at 60\;K. The design and cryogenic test results of the windows have been described briefly in~\cite{Tapio08}.

The pressure rise due to a magnet quench, at the maximum density attained during the runs with \hefour buffer gas, was limited to about 300\;mbar. During the future runs with \hethree, a quench in a closed system would result in a maximum pressure of 2.7\;bar, which requires the implementation of a safety system to limit the pressure to about 1\;bar, as will be reported in a forthcoming paper.

\begin{figure}
\begin{tabular}{cc}
\includegraphics[width=0.50\textwidth]{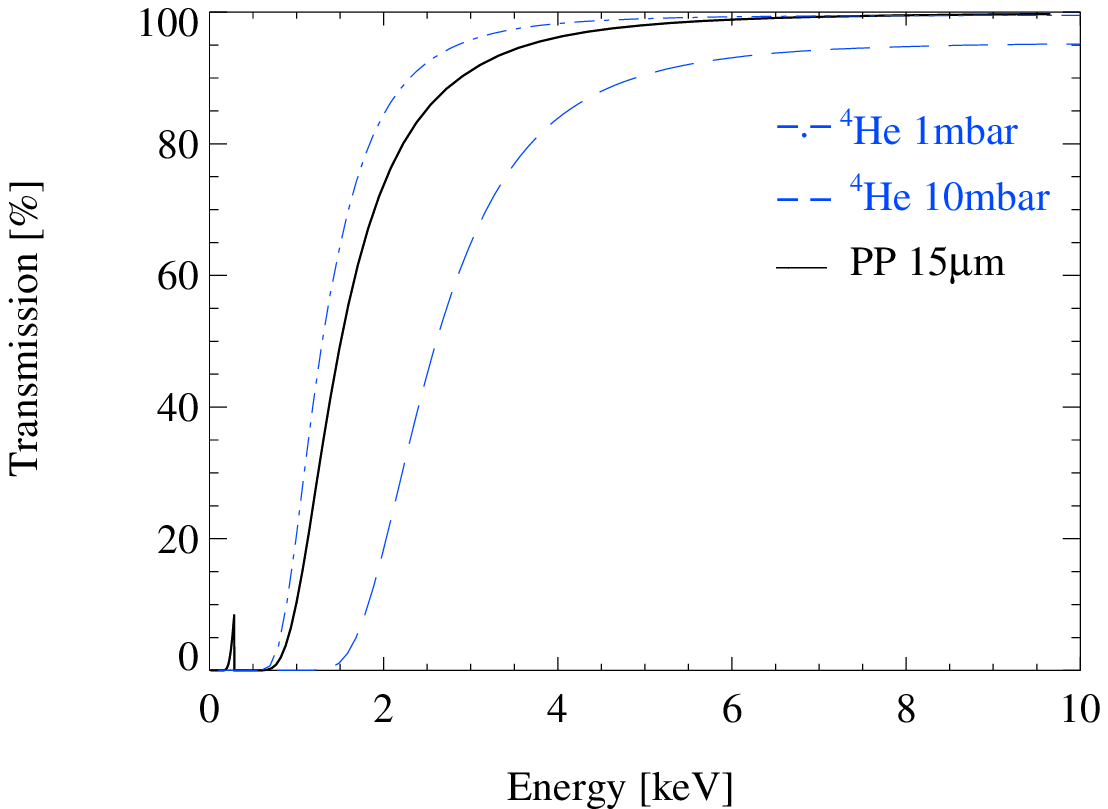}
\includegraphics[width=0.50\textwidth]{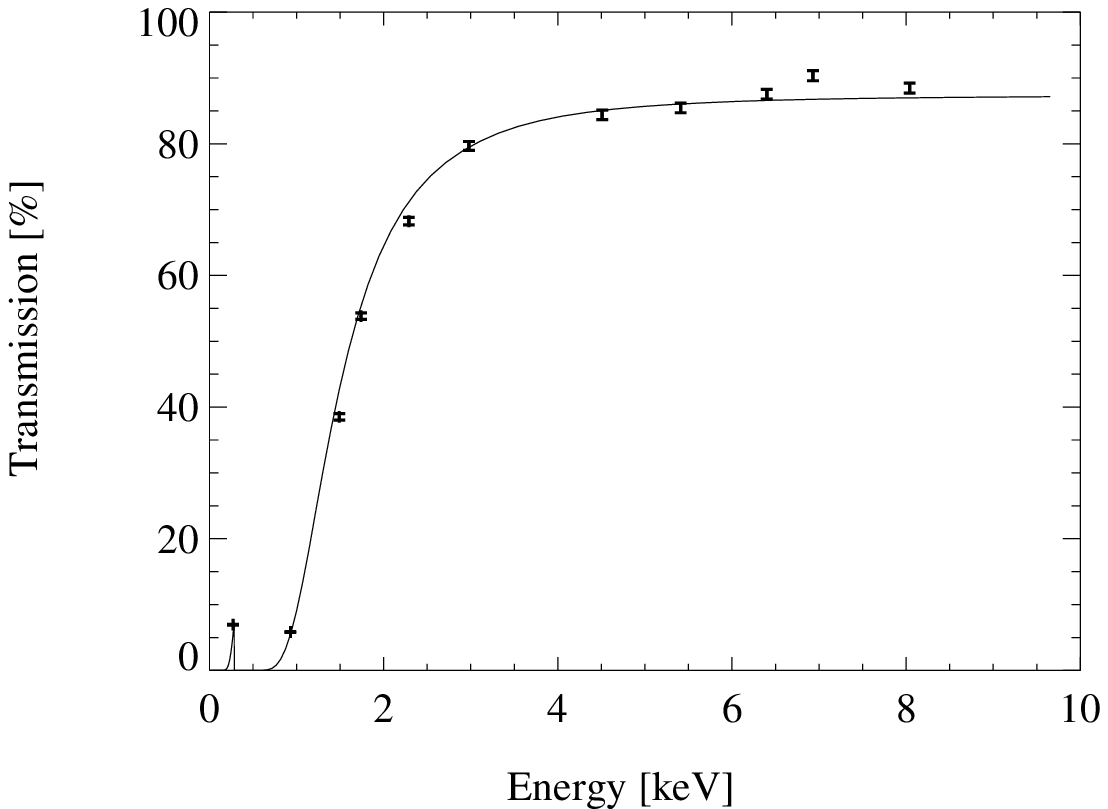} 
\end{tabular}
\caption{Left:calculated transmission of a 15\;$\mu$m film of polypropylene compared with that of 10\,m of \hefour gas at 1.8\;K and at two different pressures. Right: transmission of a 15\;$\mu$m  thick polypropylene film glued to a stainless steel strongback. The black points show the measurements performed at the PANTER x-ray facility, with errors smaller than the points. The black curve shows the calculated transmission using the NIST data and taking into account the 12.6\% transmission loss due to the strongback.}\label{fig:transmission}

\end{figure}

The transmission of the foil glued to the strongback was measured at the PANTER x-ray facility\cite{Panter}. The results of the measurements compared to a calculated transmission using NIST data and including the effect of the strongback are shown in figure~\ref{fig:transmission}. The mean calculated and measured transmission are in excellent agreement, with an overall difference of about 1\%. The absorption edge of the polypropylene foil defined the low-energy x-ray cut-off for low densities of the gas; 
however at increasingly higher densities the absorption of the gas dominates, whereas at low densities the transmission of the polypropylene dominates as shown in the left plot of figure~\ref{fig:transmission}.

The \hefour gas system was operational from November 2005 when the first \hefour runs
took place. In 2006 the data-taking period lasted for 9 months. The density was
changed in daily steps equivalent to 0.08\;mbar at 1.8\;K, covering equivalent pressures
up to 13.4\;mbar at 1.8\;K. This allowed CAST to scan a new axion mass range
between 0.02 and 0.39\;eV. During 2007, following a technical design review~\cite{He3TDR}, a more
sophisticated and complex gas system was installed and commissioned which operates
with \hethree buffer gas, and includes a pressure-limiting safety system and an improved
monitoring and control system that will enable CAST to eventually reach an axion
mass of about 1.2 eV. This new system will be described in a forthcoming paper.

\subsection{Off-resonance solar axion identification technique}

In the presence of a buffer gas which provides an effective photon mass
$m_{\gamma}$, the axion-photon conversion probability given in equation~(\ref{eq:conversionprobability}) takes the form

\begin{equation}
P_{\rm a \rightarrow \gamma}=\left( \frac{\gag\,B}{2} \right)^{2}
\frac{1}{q^{2}+\Gamma^{2}/4} \left[1+ e^{-\Gamma L} -2 e^{-\Gamma L/2}
 \cos(qL) \right],
\label{eq:probgas}
\end{equation}
\noindent where $\Gamma$ is the inverse absorption length for photons in a gas and
$q=(\ma^{2}-m_{\gamma}^{2})/(2E)$ is the axion-photon momentum
difference. In the particular case of vacuum (when $\Gamma \rightarrow 0$), equation~(\ref{eq:probgas}) becomes equation~(\ref{eq:conversionprobability}). 

\begin{figure}[h]
\begin{center}
\mbox{\epsfxsize = 8cm \epsfbox{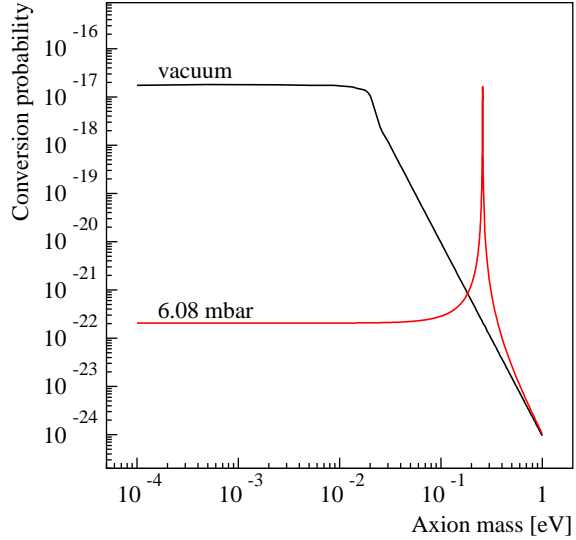}}
\end{center}
\caption{Axion-photon conversion probability versus axion mass. The black
line corresponds to  vacuum  inside the magnet pipes 
and the red line to one particular helium density setting.
Axion-photon coupling constant of $1\times 10^{-10} \mbox{ GeV}^{-1}$ is assumed.}
\label{fig:vacgas}
\end{figure}

\begin{figure}[h]
\begin{center}
\mbox{\epsfxsize = 7.6cm \epsfbox{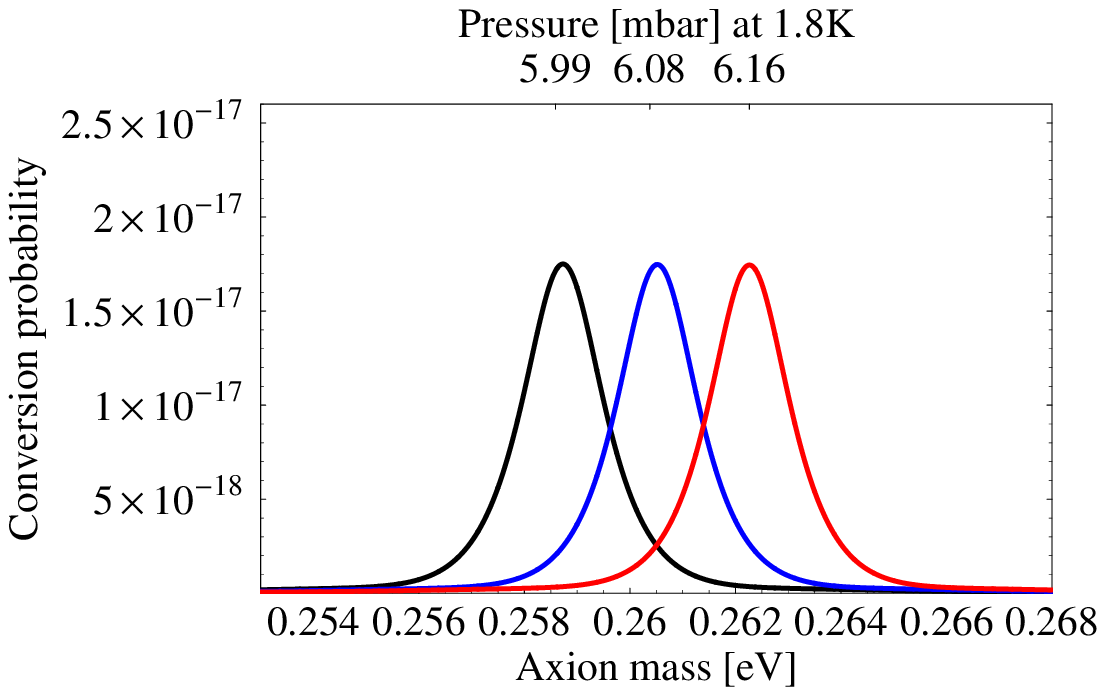}}
\mbox{\epsfxsize = 7.6cm \epsfbox{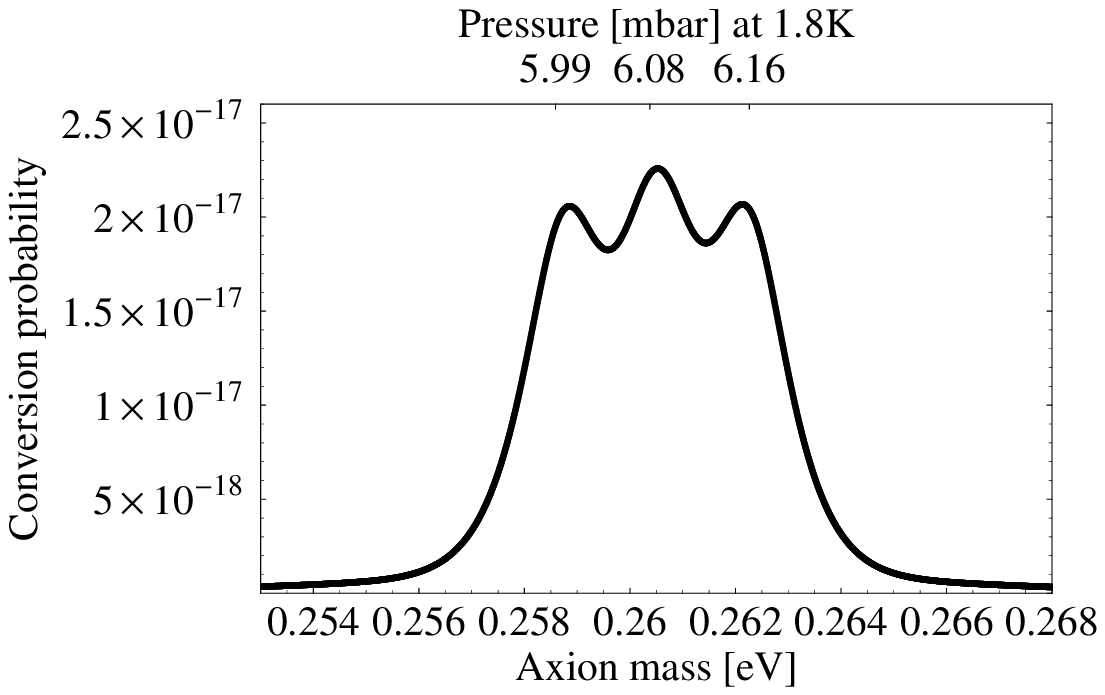}}
\end{center}
\caption{Axion-photon conversion probability versus axion mass for three
consecutive density settings on the left. On the right the same plot for the sum of the three probabilities is shown.
An axion-photon coupling constant of $1\times 10^{-10} \mbox{ GeV}^{-1}$ is assumed.}
\label{fig:3settings}
\end{figure}

The coherent axion-photon conversion will occur for $qL\lesssim\pi$, which implies
that the CAST experiment will be sensitive only to axion masses in the range
\begin{equation}
\sqrt{m_{\gamma}^{2}-\frac{2\pi E}{L}} \lesssim \ma  \lesssim
\sqrt{m_{\gamma}^{2}+\frac{2\pi E}{L}}\,\,.
\label{cohcond}
\end{equation}
For $qL\gtrsim\pi$, the axion-photon momentum mismatch will reduce the
sensitivity. During CAST-I, with vacuum inside the magnet pipes,
the coherence condition restricted the CAST sensitivity to $\ma \lesssim 0.02$\,
eV. In CAST-II, with helium inside the magnet pipes, the coherence
can be restored for a very narrow mass range ($\Delta m /m\sim 10^{-2}$--$10^{-3}$) around $m_{\gamma} \approx \ma$ (for $P =6.08\;\rm{mbar}$, for example, $\Delta m/m$ is 0.008).
The axion-photon conversion probability for two cases is shown in
figure~\ref{fig:vacgas}.
In order to cover the whole accessible axion mass range uniformly, the gas
pressure has to be varied in appropriate steps. Figure~\ref{fig:3settings} shows the
axion-photon conversion probability for three consecutive pressure settings as well as the sum of the three probabilities.
During the \hefour phase, the axion mass range $0.02 \mbox{ eV} < \ma
< 0.39 \mbox{ eV}$ was covered with 160 density settings.

\begin{figure}[ht]
\begin{center}
\mbox{\epsfxsize = 7cm \epsfbox{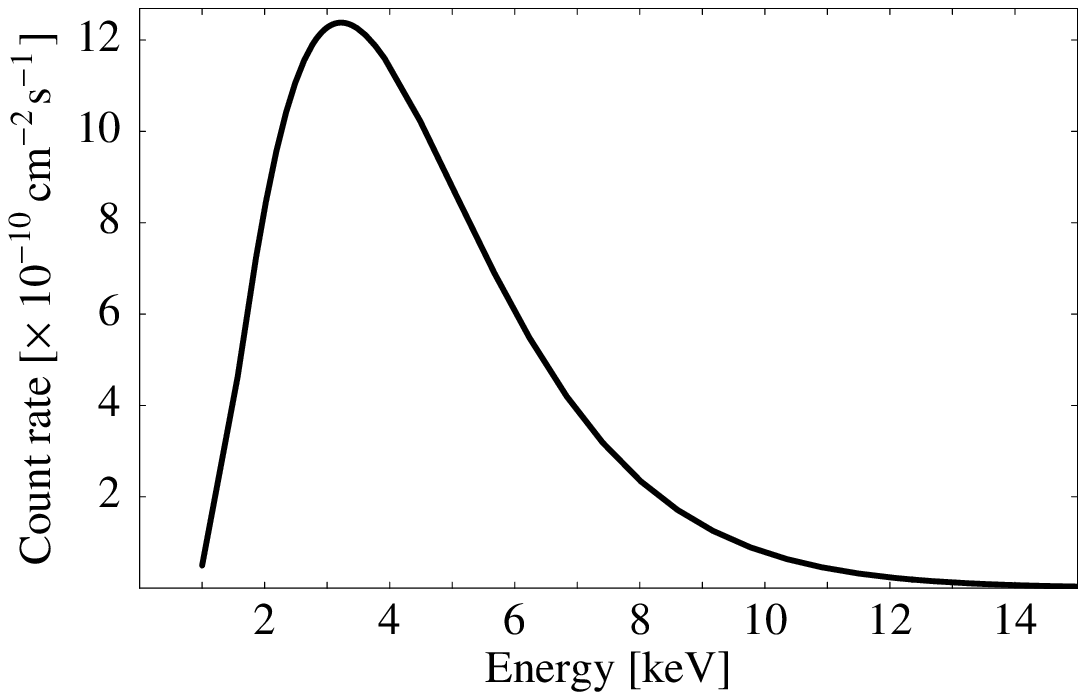}} \hspace*{0.5cm} \vspace*{0.5cm}
\mbox{\epsfxsize = 7cm
\epsfbox{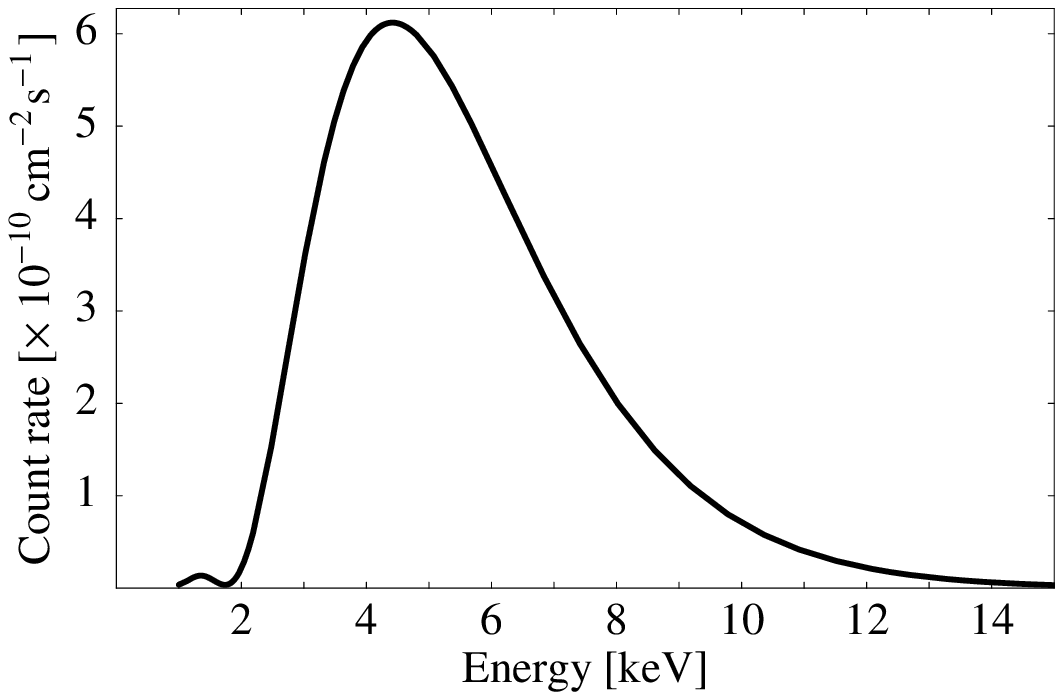}}  \mbox{\epsfxsize = 7cm \epsfbox{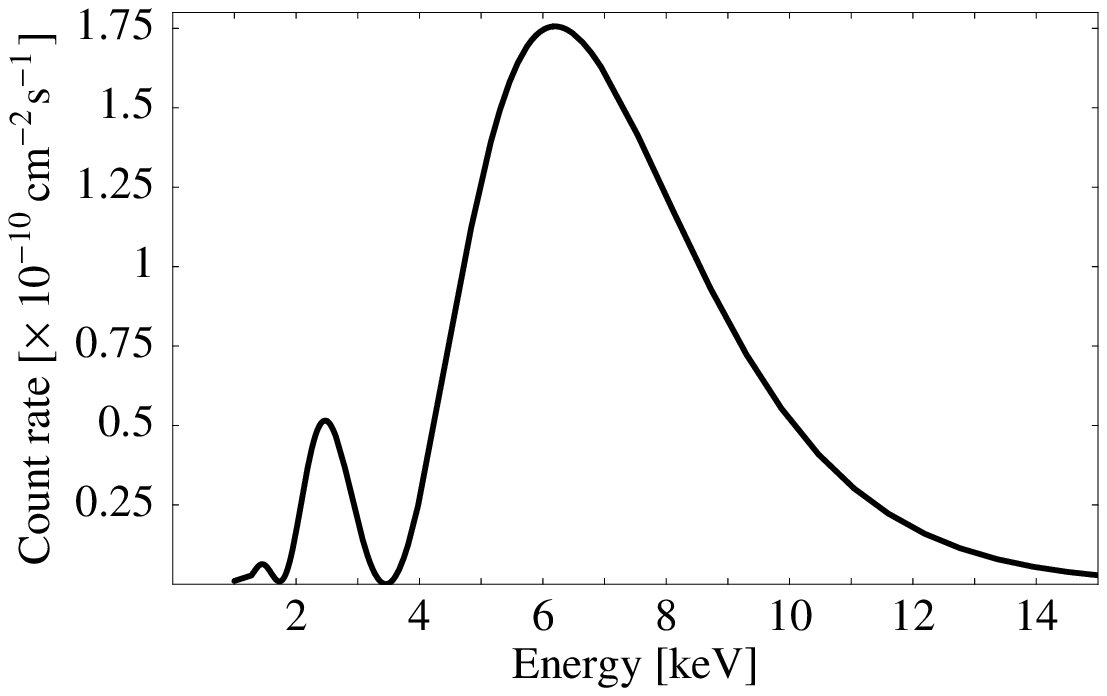}}
\hspace*{0.5cm}
\mbox{\epsfxsize = 7cm \epsfbox{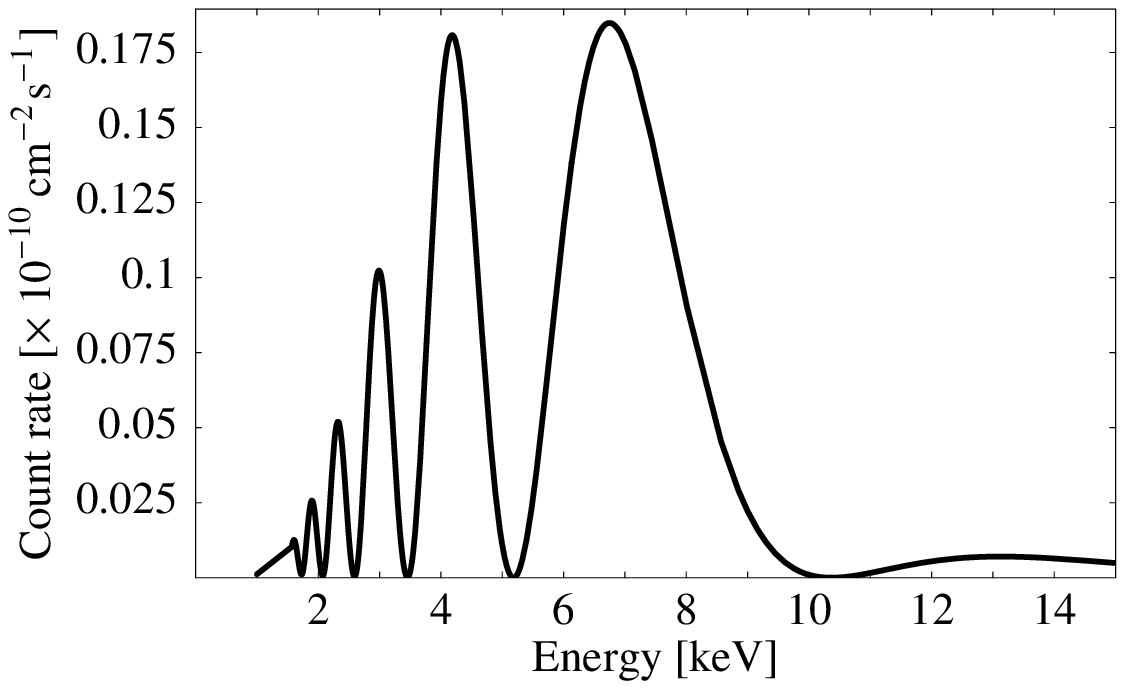}}
\end{center}
\caption{Expected photon spectra depending on the shift $S=m_{\gamma}-m_{a}$
from the resonance: $S=0$ (top left), $S=\mbox{FWHM}/2$ (top right), $S=\mbox{FWHM}$
(bottom left) $S=3 \times \mbox{FWHM}$ (bottom right). Axion-photon coupling
constant of $1 \times 10^{-10} \mbox{ GeV}^{-1}$ is assumed.}
\label{fig:offres}
\end{figure}

The solar axion flux based on the solar model in~\cite{Bahcall:2004fg} is very well approximated by (energies in keV)
\begin{equation}
\frac{d \Phi_{\rm a}}{dE}=6.02 \times 10^{10} \, g_{10}^{2} \;
 E^{2.481} \, e^{-E/1.205} \mbox{  cm}^{-2} \mbox{ s}^{-1} \mbox{ keV}^{-1},
\label{axflux}
\end{equation}
while the differential flux of photons expected at the end of the magnet in
case of the coherent conversion is given by
\begin{eqnarray}
\frac{d \Phi_{\gamma}}{dE} & = & \frac{d \Phi_{\rm a}}{dE} \, P_{\rm a \rightarrow \gamma}
\\
 & = & 0.088 \; g_{10}^{4}
\left( \frac{L}{9.26 \mbox{ m}} \right)^{2} \left( \frac{B}{9.0 \mbox{ T}} \right)^{2}
E^{2.481} \, e^{-E/1.205} \mbox{  day}^{-1} \mbox{ cm}^{-2} \mbox{ keV}^{-1},
\nonumber
\label{phflux}
\end{eqnarray}
where $g_{10} = g_{\rm a \gamma} / (10^{-10} \mbox{ GeV}^{-1})$.
CAST developed a novel technique for solar axion identification using the off-resonance spectral distribution. The expected photon spectrum depends on the resonance mismatch between \ma~and $m_{\gamma}$ as shown in figure~\ref{fig:offres}. The characteristic spectral distribution can be used for additional confirmation of the signal as a solar axion. We remark that this possibility, to the best of our knowledge, was never used before in axion search experiments.

\section{Data analysis}\label{sec:analysis}
During the data-taking period with  \hefour in the magnet bores in 2005 and 2006, the x-ray detectors were operated
in the same configuration as for the 2004 data taking period~\cite{And07,Kus07,Aut07,Abb07}, except for minor
improvements. The time projection chamber (TPC)~\cite{Aut07}, covering both bores of the east end of the magnet looking
for axion-to-photon conversion during sunset, had reduced electronic noise, using improved low voltage
power supplies. At the other end, looking for x-rays from axion-to-photon conversion during sunrise, two detection
systems were installed: a gaseous micromegas chamber (MM)~\cite{Abb07} and an x-ray telescope consisting of x-ray optics coupled to a pn-Charge Coupled Device (CCD)~\cite{Kus07} as a focal plane detector. 
The use of the x-ray mirror system suppresses the background by a factor of about 155 since the potential signal from the magnet acceptance area of 14.5~$\mathrm{cm}^2$ is focused to a spot of roughly 9.3~$\mathrm{mm}^2$ on the CCD chip, thus improving the signal to background ratio by the same factor.
The vacuum system of the telescope and the CCD was upgraded as well as the control software
resulting in a safer and more flexible mode of operation. 
A new MM detector was installed for Phase II which had an improved performance with respect to the one used previously: a reduction of the copper fluorescence due to detector materials, which used to dominate the detector background, was achieved by the introduction of a gold-coated amplification mesh.

As during CAST-I, axion-sensitive data were taken when the magnet was pointing to the sun (about
$2\times 1.5$\;h per day), while the rest of the time was used to measure background continuously only interrupted by daily calibrations. 
The density setting was changed once a day so that both sunrise and sunset detectors could cover all settings. Small shutdown periods, for a single detector, due to replacements of components, maintenance or upgrades did occur. As a general rule, the pressure step was repeated if more than one of the detectors were off. 
\begin{figure}
\begin{tabular}{cc}
\includegraphics[width=0.33\textwidth,angle=-90]{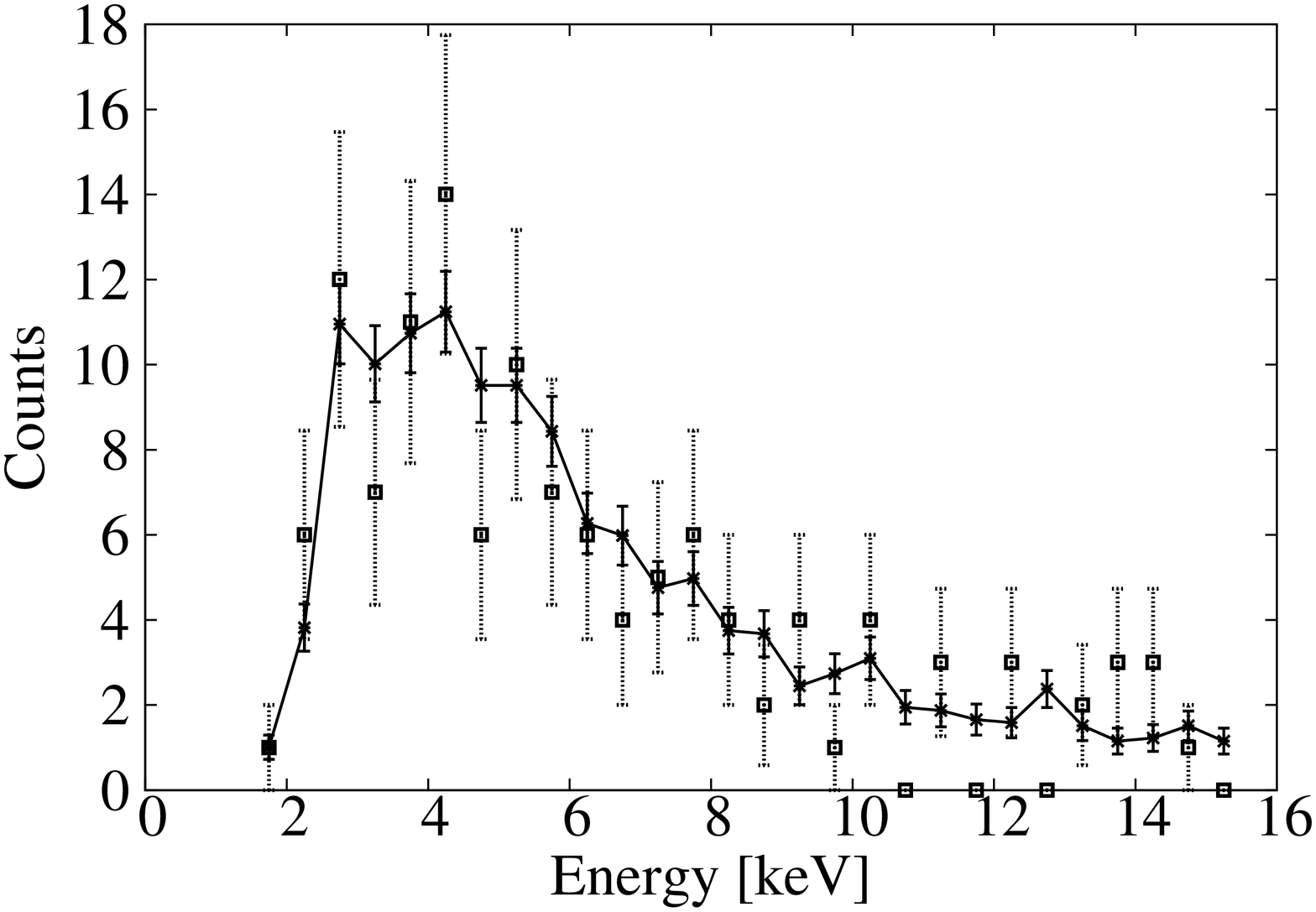} &
\includegraphics[width=0.33\textwidth,angle=-90]{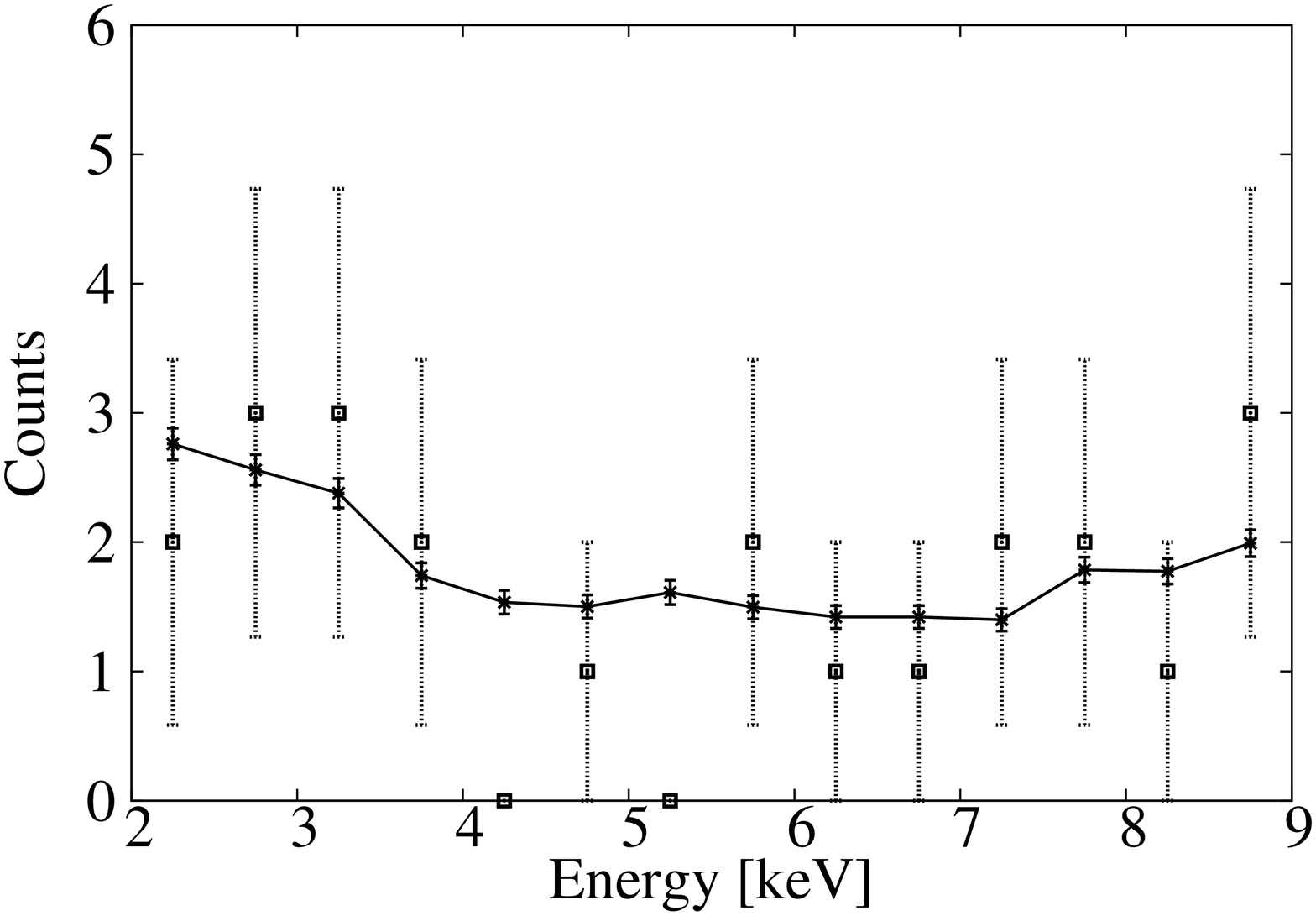} 
\end{tabular}
\caption{ Energy distribution of events recorded
    during the tracking run (stars with dashed line) at pressure setting
    $P_{k}=8.909\,\text{mbar}$ compared to background data (empty squares with continuous line) for the TPC (left) and the Micromegas (right) detectors respectively.}\label{fig:MMTPC_spectra}
\end{figure}

\begin{figure*}
  \centering
  \begin{minipage}{0.49\textwidth}
    \centerline{\includegraphics[height=0.55\textheight,angle=0]
      {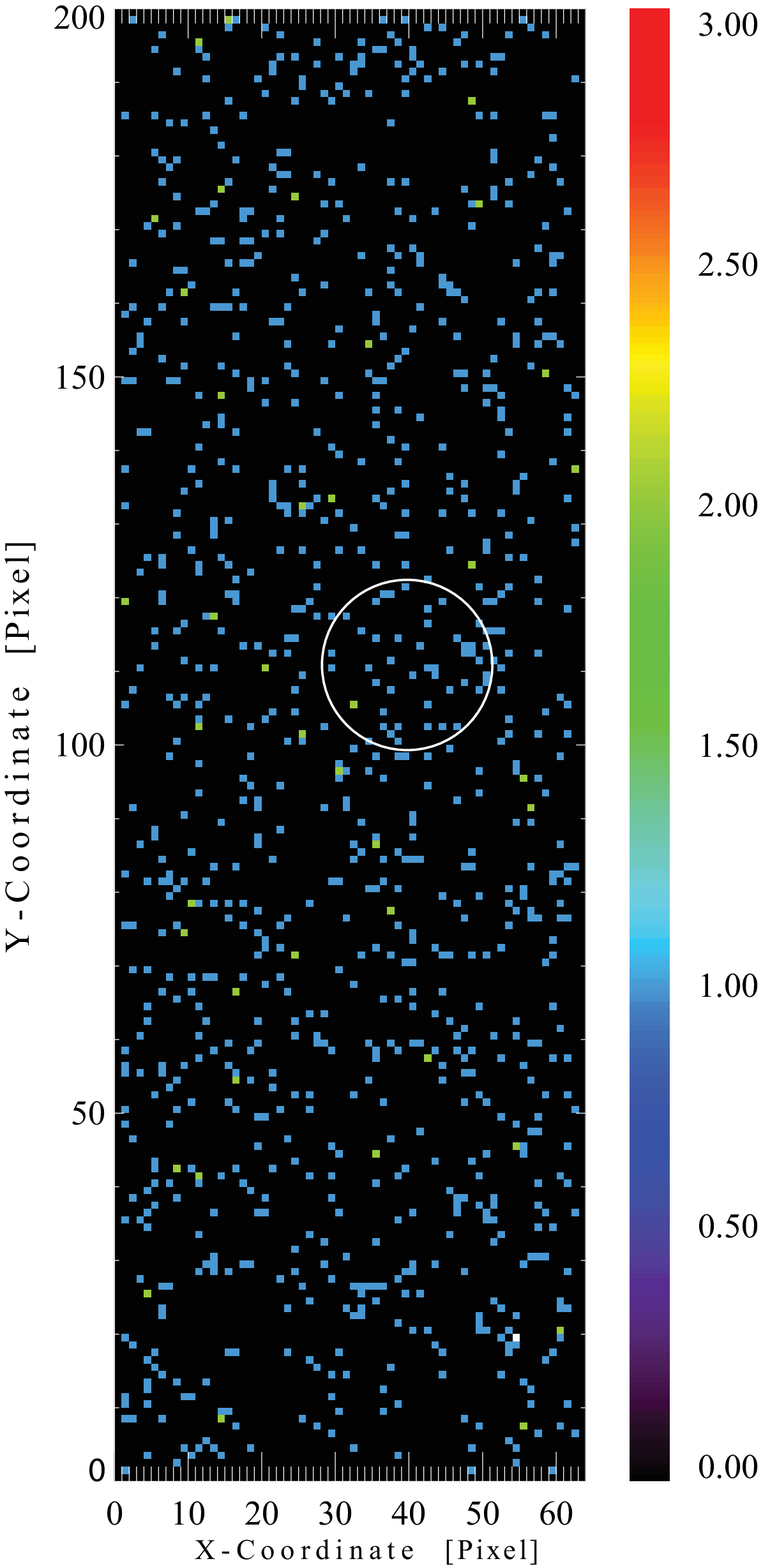}}
  \end{minipage}
  \begin{minipage}{0.49\textwidth}
    \centerline{\includegraphics[height=0.55\textheight,angle=0]
     {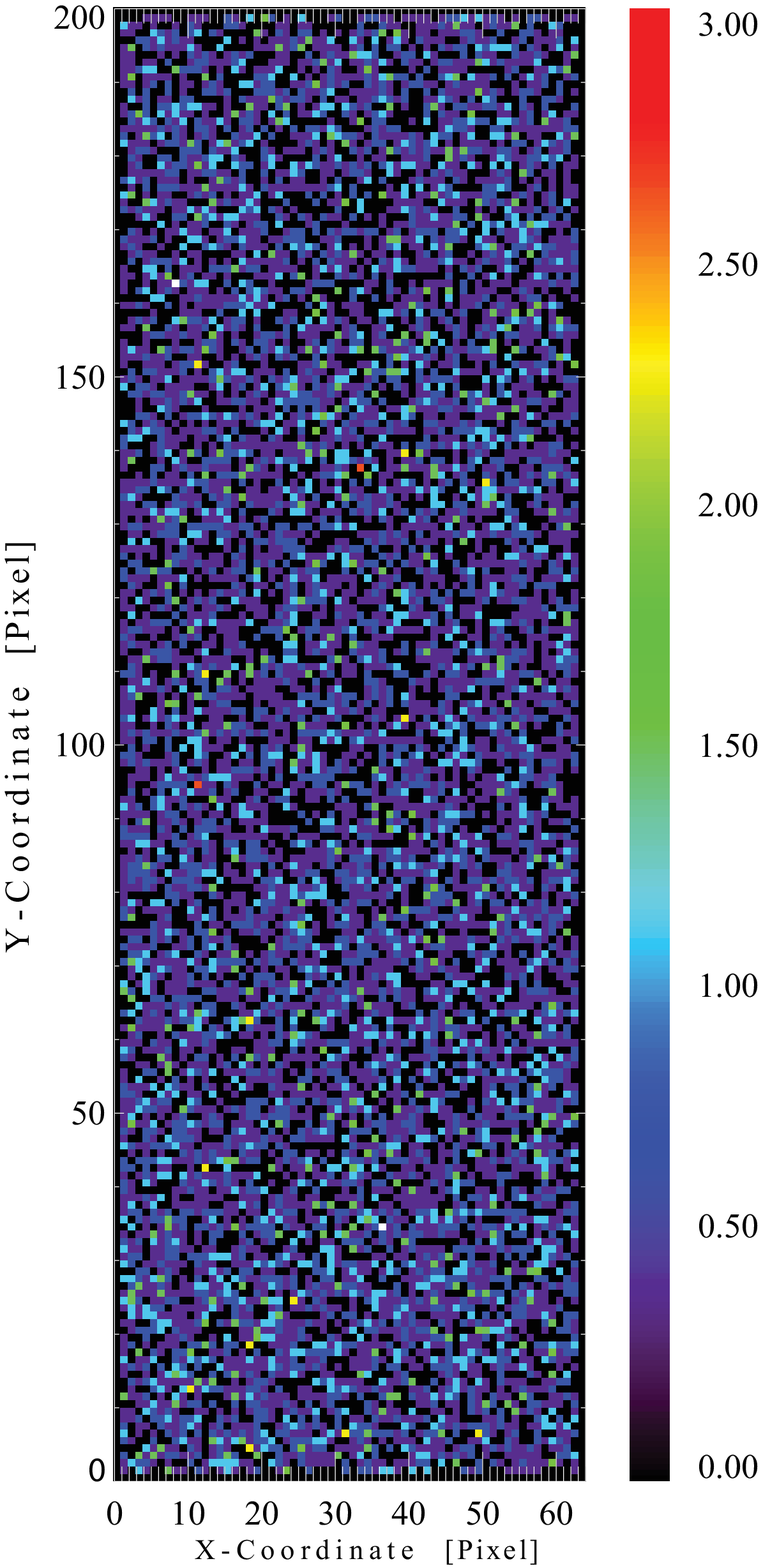}}
  \end{minipage}
  \caption{\label{fig:CCD_image}Left: Spatial
    distribution of events observed during sun tracking by the
    CAST x-ray telescope during the 2005/2006 data taking period (189 solar
    observations). The intensity is given in counts per pixel and is
    integrated over the tracking period of $294.8\;\text{h}$ accounting to 45 counts between 1 and 7\,keV. The white circle represents the 
    expected size and position of the potential signal, which would be an image of the sun's hot inner axion-producing region. 
    Right: Background spatial distribution as observed by the CAST x-ray telescope
    during the 2005/2006 data-taking period. The intensity is given in
    counts per pixel and integrated over the full observation period of
    $2741.5\;\text{h}$.}
\end{figure*}

An example of the energy distribution of the events during the tracking run at the density setting $P_{k}=8.909\,\text{mbar}$ at 1.8\;K compared to background data are shown in figure~\ref{fig:MMTPC_spectra} for the TPC and MM detectors. The background data shown in the plots is the effective background used to calculate the axion-photon coupling constant. Figure~\ref{fig:CCD_image} shows the background data observed in the CCD detector (area $3\times1\,$cm$^2$) during the \hefour~run. The solar axion signal is expected to cover a circular spot of $\approx9.3\,\mathrm{mm}^2$ and is indicated in figure~\ref{fig:CCD_image} by the white circle. 
The stability of the alignment of the optical axis of the x-ray telescope to the magnet axis has been monitored with an external x-ray source to an accuracy of $\pm 0.5$~pixel. Therefore, the location of the potential signal spot on the CCD chip is known with the same precision throughout the run.
A summary of the data acquired with each of the detectors for both solar tracking and background
data is given in table \ref{tab:data04}. In order to extract the final result the data from the three detectors are  combined.
\begin{table*}[h]
  \centering
  \footnotesize
\caption{\label{tab:data04} Summary of data taken during the \hefour
phase. Average background rates  are given for each detector. }
  \begin{tabular}{cccccccc} \hline

    & \multicolumn{1}{c}{Steps} & \multicolumn{1}{c}{Tracking} &
    \multicolumn{1}{c}{Background} &
    \multicolumn{1}{c}{Rate}  &
    \multicolumn{1}{c}{Energy Range}  \\
    & & (h) & (h) &${\rm keV}^{-1}{\rm s}^{-1}{\rm cm}^{-2}$ & keV\\ \hline
     CCD  & 147   & 294.8       & 2758.1 &  $(8.66 \pm 0.06)\times10^{-5}$& 1--7 \\
     TPC  & 154   & 304.1       & 4346.6 &  $(7.68 \pm 0.01)\times10^{-5}$& 2--15 \\
     MM   & 159   & 336.6       & 3115.0 &  $(4.75 \pm 0.02)\times10^{-5}$& 2--9 \\ \hline
  \end{tabular}
\end{table*}

During Phase I, the presence of a solar axion would have been evident over the
entire data taking period of roughly one year. 

The signal to be looked for in the CAST-II data
would be present only in a few trackings centered at the density which matches the axion mass observation corresponding to one density setting. The effective exposure time at a given axion mass  is approximately 100\;minutes, during which a very low number of counts is expected as background (e.g. for the x-ray telescope only about 0.26\;counts are expected in the spot area during one tracking). Hence, the sensitivity of CAST in phase II is statistics-limited by the number of expected background  events.

In order to extract an axion signal or to derive an upper limit on its coupling constant from CAST-II data, one has to take into account
the fact that the axion signal $s$ depends on the density at step $k$ at which data are
taken, $s = s_k(g_{\rm a\gamma} , \ma)$, and that $s_k$ is maximum when $\ma$ matches the gas-induced
photon mass $m_k$,
$m_k^2 = 4\pi \alpha n_{ek}/m_{e}$, $n_{ek}$ being the electron density at setting $k$, 
but quickly drops as $m_k$ deviates from $\ma$. A standard likelihood function can be built for a single step $k$ based on the Poissonian probability distribution,
\begin{equation}
\label{singleL}
     \mathcal{L}_k(\gag,\ma) = \frac{P_k}{P_{0k}} = \frac{P(\{n_i\}_k;\{\mu _i\}_k)}{P(\{n_i\}_k;\{n_i\}_k)},
\end{equation}
\noindent where \\[3pt]
\begin{equation}
P(\{n_i\}_k;\{\mu _i\}_k)= \prod_{i} e^{-\mu _{ik}} \frac{\mu_{ik}^{n_{ik}}}{n_{ik}!} \; ,
\end{equation}

\noindent $n_{ik}$ is the observed number of counts in energy bin $i$ and density step $k$, and $\mu_{ik}$ is the expected
number of counts in each bin, estimated as the sum of expected background
counts plus the axion signal $\mu_{ik}=b_{ik}+s_{ik}(\gag,m_{\rm a})$. The estimation of
$b_{ik}$ is done using experimental data taken in non-tracking conditions, following different prescriptions
that are discussed below. The statistical uncertainty in $b_{ik}$ is not taken into account in equation (\ref{singleL}) since  its effect on the final result has been shown to be negligible by a dedicated Monte-Carlo simulation.

Data from different density steps can be combined by
multiplying the corresponding $\mathcal{L}_k$,
\begin{equation}\label{combinedL}
     \mathcal{L}(\gag,\ma) = \prod_k \mathcal{L}_k(\gag,\ma).
\end{equation}

\noindent The use of $\mathcal{L}$ to extract statistical information, like best-fit values or confidence intervals, provides a method
to account for neighbouring pressure steps and off-resonance axion masses. The product
in equation (\ref{combinedL}) runs in principle over all $k$ settings,
in practice, however, only those settings close to the axion mass \ma~evaluated contribute to  the final likelihood
function, a fact used to reduce computation time. In order to combine the data from the three x-ray detectors, the  calculated $\mathcal{L}$ in equation~(\ref{combinedL}) for individual detectors  are multiplied.

\begin{figure}[htb!]
\centering
\begin{tabular}{c}
\includegraphics[width=0.55\textwidth]{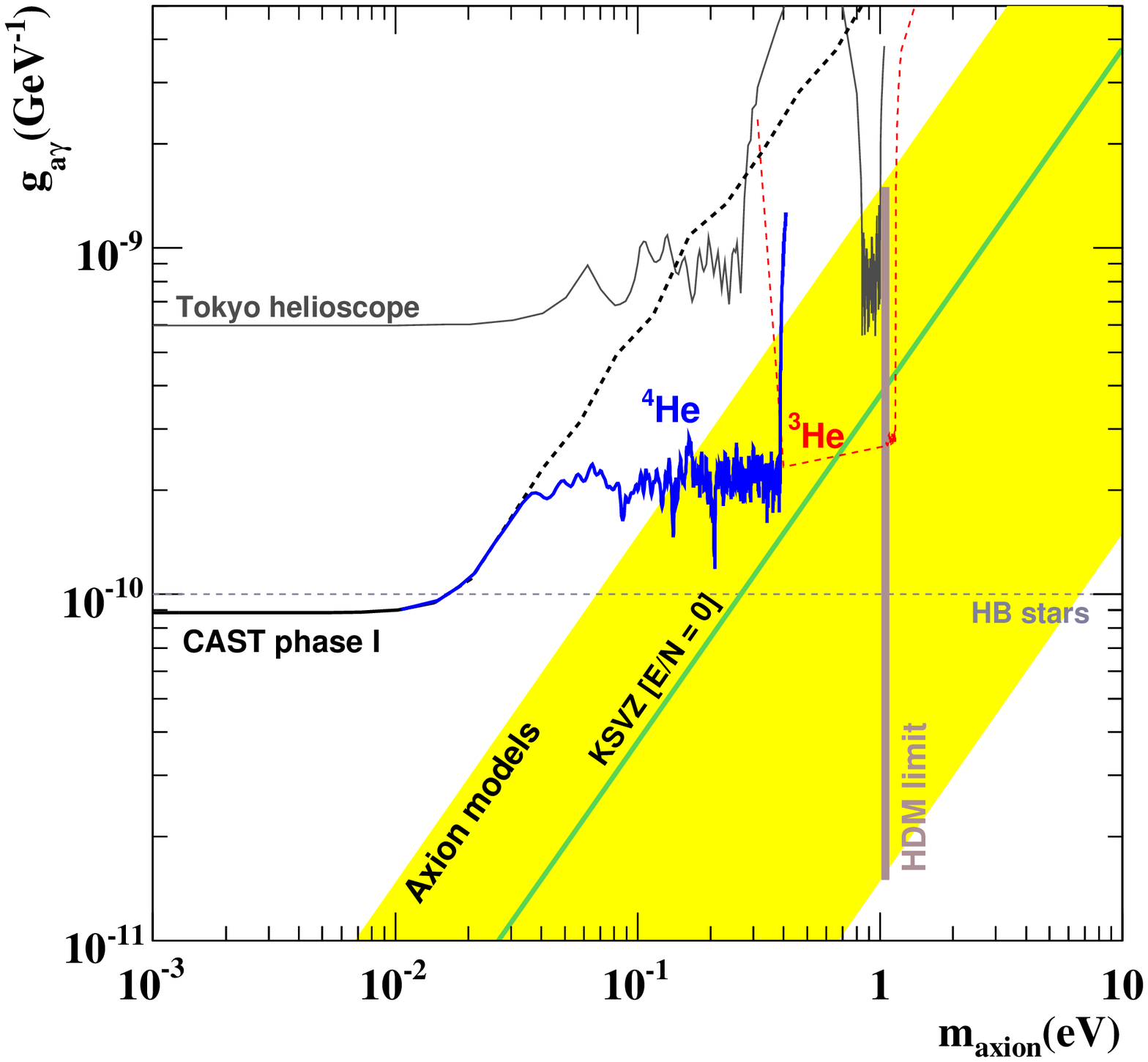} \\
\includegraphics[width=0.55\textwidth]{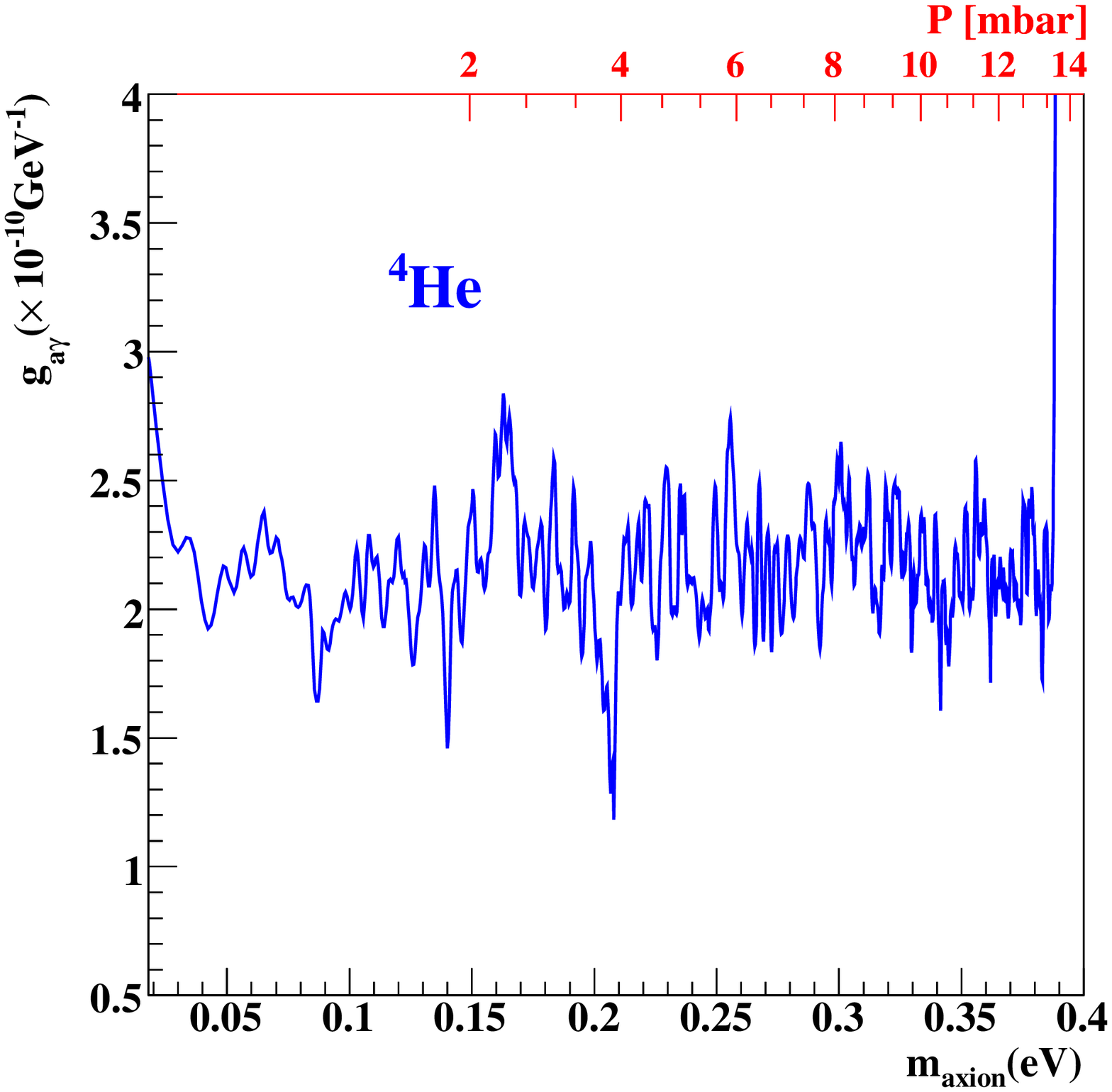}
\end{tabular}
\caption{ Top: Exclusion plot in the axion-photon coupling versus the axion
   mass plane. The limit achieved by the CAST experiment (combined result of the CAST-I
   and \hefour part of CAST-II) is compared with constraints from the
   Tokyo helioscope~\cite{Moriyama:1998kd,Inoue:2002qy,Inoue:2008} and HB
   stars~\cite{Raffelt:2006cw,Raf96})
   discussed in the Introduction. The vertical line (HDM) is the
   hot dark matter limit for hadronic axions
   $\ma<1.0\;{\rm eV}$ ~\cite{Hannestad:2005df,Hannestad:2008js} inferred
   from observations of the cosmological large-scale structure.
   The yellow band represents typical theoretical models with
   $\left|E/N-1.95\right|$ in the range 0.07--7 while the green solid line
   corresponds to the case when $E/N=0$ is assumed. The red dashed line shows our prospects for the \hethree run started in March 2008. Bottom: Expanded view of the limit achieved in the \hefour part CAST-II for \ma~between 0.02\;eV and 0.39\;eV corresponding to a pressure scan from 0 to 13.4\;mbar.}\label{fig:exclusion}
\end{figure}

The final step is now similar to the one followed in our previous analysis~\cite{And07}. A best-fit value $g_{min}^4$ is obtained after maximisation of $\mathcal{L}$ for a fixed value of \ma. The results are compatible with absence of signal, and therefore we can express our result as an upper limit on the axion photon coupling with a 95\% confidence level, $g_{95}^4$. To do that we follow the Bayesian approach, which consists in considering the Bayesian probability $P(g^4|g^{4}_{min}$)  with a prior distribution uniform in $g^4$ 
and integrating it from zero to 95\% of its area, in order to find $g_{95}^4$.
This value is computed for
many values of the axion mass \ma~in order to configure the full
exclusion plot shown in figure~\ref{fig:exclusion}.  This plot shows the combined CAST-I and \hefour part of CAST-II
results (blue line) along with the constraints from the  Tokyo helioscope~\cite{Moriyama:1998kd,Inoue:2002qy,Inoue:2008} and HB stars~\cite{Raffelt:2006cw,Raf96}. The vertical line (HDM) is the Hot Dark Matter limit for hadronic axions ($\ma<1.0\;{\rm eV}$) ~\cite{Hannestad:2005df,Hannestad:2008js} inferred from observations of the cosmological large-scale structure. The yellow band represents typical theoretical models with $\left|E/N-1.95\right|$ in the range 0.07--7 where the green solid line corresponds to the case $E/N=0$. The red dashed line shows our prospects for the \hethree run started in March 2008. The plot shows an increase of sensitivity at discrete masses (e.g. $\ma\sim0.2\;$eV corresponding to a pressure $P_{k}= 3.747\;$mbar at 1.8\;K). This reflects the fact that more time was spent at these pressure settings in order to follow what resulted to be statistical fluctuations. A close-up of the same plot showing the axion mass range explored
in the \hefour~part of CAST-II is also shown.

The influence of systematic uncertainties on the best-fit value of $g^4$ and on the upper limit of \gag~has been studied for all experimental parameters entering the analysis: magnet length and field, detector efficiencies, window transparencies, etc. Most of these sources of uncertainty have negligible effects on the final result (less than 1\%), the exception being the background definitions and, for  the x-ray telescope, the tracking precision. 
 
Any pointing inaccuracy affects negatively the effective area of the telescope and the expected location and size of the signal spot on the CCD. For our stated accuracy of 0.01\;degree, these effects have been quantified to introduce a systematical uncertainty in the final limit of less than 3\%.
 
Regarding the background determination, different prescriptions have been used in order to estimate the systematic uncertainty in our ability to measure the background of each detector for each density step, induced by possible uncontrolled dependencies of the background on time, position or other experimental conditions. 
 
For the TPC detector, the background $b_{ik}$ to be used for a specific density step $k$ is usually defined using data taken in the non-tracking runs of the same day(s) (i.e. same pressure in the magnet) as the tracking run of that step. Alternatively, data from neighboring days (up to 5 days after and before) have also been used. Another background definition makes use of the tracking data from off-coherence pressure steps. For the MM detector, in order to check for possible diurnal effects, background measurements taken at different daytime have been compared. For the case of the CCD, the data outside the solar spot, taken either during tracking or non-tracking runs have been used alternatively or in combination with the data from the spot area.
 
The variation of the final result due to the use of these different background prescriptions is usually much smaller than 10\% and only in extreme cases of that order. Therefore we estimate that the overall effect of systematic uncertainties on our final combined upper limit of \gag~ is less than 10\%.
 
As can be seen in figure~\ref{fig:exclusion}, CAST extends its exclusion line from axion masses of 0.02\;eV (Phase I) up to masses of 0.39\;eV. 
For the \hefour~phase the limit at a given mass is derived from a few hours of data taking only and correspondingly small event numbers causing large statistical fluctuations of the line contour.

For the first time, limits resulting from direct observation have entered the QCD axion model band in the eV range, excluding an important part of the parameter space.

\section{Conclusions}                              \label{sec:conclusion}

The CAST-I search for solar axions has provided the most restrictive
observational limit on the two-photon coupling of axions and axion-like
particles (ALPs) for $\ma\lesssim0.02$\;eV~\cite{And07}.  In the
first part of CAST-II setup, we have used $^4$He as a buffer gas to provide
x-rays with an effective mass within the magnet bores.  Varying the gas
density in 160 steps, we have extended the search up to $\ma\lesssim0.4$\;eV. 
The absence of a signal above background excludes a new range in the \gag--\ma~plane 
shown in figure~\ref{fig:exclusion} that was not previously explored by direct laboratory experiments.
We were able to derive a mean upper limit of $2.17 \times 10^{-10}\;\text{GeV}^{-1}$ in the range $0.02 < \ma < 0.39$ eV.

The CAST search has now entered the realistic model parameter space
for QCD axions. In contrast to generic axion-like particles, QCD
axions unavoidably interact with nucleons so that one expects them to
be efficiently emitted from a hot nuclear medium. Accordingly, the
well-known energy-loss argument based on the duration of the observed
neutrino burst of SN~1987A provides a limit corresponding roughly to
$m_{\rm a}\lesssim 10^{-2}$~eV. A concise and recent  summary of the status of this bound can be found 
in Sec. 6 of \cite{Raf08} and refs. therein. The SN~1987A limit is a powerful argument
that has been applied to many cases other than axions, but on the
other hand it suffers from the very sparse statistics of the SN~1987A
neutrino burst as well as possibly large systematic uncertainties from
the axion emission rates in dense nuclear matter. For sure a CAST
detection of axions in the range $m_{\rm a}\gtrsim10^{-2}$~eV 
would reveal a significant problem with the often-used SN 1987A argument.

In the ongoing second part of CAST-II, we use \hethree as a buffer
gas, allowing us to reach higher gas pressures at the operating
temperature of 1.8~K and thus to reach axion masses up to about
$1~\text{eV}$.  In this mass range QCD axions would exist as thermal
relic particles in the universe and provide a hot dark matter
component similar to neutrinos~\cite{Moroi:1998qs}. Therefore, the
usual structure-formation arguments that provide neutrino mass limits
can be applied to axions as
well~\cite{Hannestad:2005df,Melchiorri:2007cd,Hannestad:2007dd,Hannestad:2008js}.
Based on the latest cosmological data that are safely in the linear
regime of structure formation, a limit on the axion mass of $m_{\rm
  a}<1.0$~eV at 95\% CL is found~\cite{Hannestad:2008js}.  In the long
run such cosmological results are likely to improve, with the ultimate
goal of detecting the unavoidable neutrino hot dark matter
component. If eventually a hot dark matter component above the minimal
neutrino contribution is found in cosmological precision data, its
interpretation is not necessarily unique and could signify an axion
component. In such a case, and if the forthcoming CAST search is
unsuccessful, the experimental challenge for axion searches in the
sub-eV range is to improve the sensitivity beyond the stellar
evolution limits and to cover the full range of plausible axion models
that is roughly represented by the yellow band in
figure~\ref{fig:exclusion}.

\section*{Acknowledgments}
CAST wishes to give special thanks to the family Poulin-McGilchrist for kindly granting permission to remove three oak trees which were in line of sight for the sun filming tests carried out each March and September and which are considered an essential crosscheck of the alignment of our experiment.  
We thank CERN for hosting the experiment and for the contributions of V. Benda, J. P. Bojon, F. Cataneo, R. Campagnolo, G. Cipolla, F. Chiusano, M. Delattre, L. Dufay-Chanat, J.-F. Ecarnot, F. Formenti, D. Fraissard, M. Genet, J. N. Joux, L. Le Mao, A. Lippitsch, L. Musa, R. De Oliveira, A. Onnela, J. Pierlot, S. Prunet, C. Rosset, H. Thiesen, A. Vacca, and B. Vullierme.
We acknowledge support from MSES (Croatia) under grant No. 098-0982887-2872, CEA (France), BMBF (Germany) under grant Nos. 05 CC2EEA/9 and 05 CC1RD1/0, the Deutsche Forschungsgemeinschaft (DFG) under grant HO 400/7-1, GSRT (Greece), RFFR (Russia), the Spanish Ministry of Science and Education (MEC) under grants FPA2004-00973 and FPA2007-62833 and the Turkish Atomic Energy Authority. This work was supported in part by the U.S. Department of Energy under Contract No. DE-AC52-07NA27344; support from the Laboratory Directed Research and Development Program at LLNL is also warmly acknowledged.
We acknowledge helpful discussions within the European Union network on direct dark matter detection of the ILIAS integrating activity (Contract number: RII3-CT-2003-506222). 

The CAST collaboration dedicates this paper to the memory of our colleagues E. Arik, F. S. Boydag, O. B. Dogan and I. Hikmet, who perished in the Atlasjet accident in southern Turkey, 30 November 2007. Their friendship, energy and enthusiasm will be sorely missed.

\section*{References}

\end{document}